\documentclass[reprint,aps,prd,superscriptaddress,preprintnumbers,nofootinbib]{revtex4-1}

\usepackage{amsfonts}
\usepackage{amsmath}
\usepackage{amsthm}
\usepackage{amssymb}
\usepackage{array}
\usepackage{dcolumn}
\usepackage[svgnames]{xcolor}
\usepackage[hidelinks]{hyperref}
\hypersetup{
    colorlinks=true,
    linkcolor=NavyBlue,
    filecolor=NavyBlue,
    urlcolor=NavyBlue,
    citecolor=NavyBlue,
    }

\usepackage{graphics}
\usepackage{tikz}
\usepackage{graphicx}
\usetikzlibrary{positioning}
\usepackage{caption}
\usepackage{subcaption}
\usepackage{adjustbox}
\usepackage{gensymb}
\usepackage{breqn}
\usepackage{braket}
\usepackage{enumitem}
\usepackage{gensymb}

\def\be{\begin{equation}}
\def\ee{\end{equation}}
\def\bea{\begin{eqnarray}}
\def\eea{\end{eqnarray}}

\def\gev{\, {\rm GeV}}
\def\mev{\, {\rm MeV}}
\def\kev{\, {\rm keV}}
\def\ev{\, {\rm eV}}

\newcommand{\gsim}{\lower.7ex\hbox{$\;\stackrel{\textstyle>}{\sim}\;$}}
\newcommand{\lsim}{\lower.7ex\hbox{$\;\stackrel{\textstyle<}{\sim}\;$}}

\begin{document}

\hfill \preprint{MI-HET-785, UH511-1329-2022}

\title{Low-Mass dark matter (in)direct detection with inelastic scattering}

\author{Nicole F.~Bell}
\email{n.bell@unimelb.edu.au}
\affiliation{ARC Centre of Excellence for Dark Matter Particle Physics$,$ \\~School of Physics$,$~ The~ University~ of~ Melbourne$,$~ Victoria~ 3010$,$~ Australia}

\author{James B.~Dent} 
\email{jbdent@shsu.edu}
\affiliation{Department of Physics$,$~ Sam ~Houston~ State~ University$,$~ Huntsville$,$~ TX~ 77341$,$~ USA}

\author{Bhaskar Dutta}
\email{dutta@physics.tamu.edu}
\affiliation{Mitchell Institute for Fundamental Physics and Astronomy$,$ Department~ of ~Physics ~ and~ Astronomy$,$\\ Texas A$\&$M University$,$~College~ Station$,$ ~Texas ~77843$,$~ USA}

\author{Jason Kumar}
\email{jkumar@hawaii.edu}
\affiliation{Department of Physics and Astronomy$,$~ University~ of~ Hawaii$,$~ Honolulu$,$~ Hawaii~ 96822$,$~ USA}

\author{Jayden L.~Newstead}
\email{jnewstead@unimelb.edu.au}
\affiliation{ARC Centre of Excellence for Dark Matter Particle Physics$,$ \\~School of Physics$,$~ The~ University~ of~ Melbourne$,$~ Victoria~ 3010$,$~ Australia}

\begin{abstract}
   We revisit the detection of luminous dark matter in direct detection experiments. In this scenario, dark matter scatters endothermically to produce an excited state, which decays to produce a photon. We explore ways in which the electron recoil signal from the decay photon can be differentiated from other potential electron recoil signals with a narrow spectral shape. We find that larger volume/exposure xenon detectors will be unable to differentiate the signal origin without significant improvements in detector energy resolution of around an order of magnitude. We also explore what can be learned about a generic luminous dark matter signal with a higher resolution detector. Motivated by the advancements in energy resolution by solid-state detectors, we find that sub-eV resolution enables the discovery of LDM in the presence of background levels that would otherwise make observation impossible. We also find that sub-eV resolution can be used to determine the shape of the luminous dark matter decay spectrum and thus constrain the dark matter mass and velocity distribution.
\end{abstract}

\maketitle
\section{Introduction} \label{sec:introduction}

It is a well-established fact that we cannot account for all of the gravitational mass with observable baryonic matter. The missing mass, or dark matter, can be observed from subgalactic to cosmological length scales~\cite{deSalas:2020hbh,Planck:2018vyg}. While it is common to solve the dark matter problem through the introduction of a new weakly coupled particle, there are a variety of scenarios in which the dark sector consists of multiple particles~\cite{TuckerSmith:2001hy, TuckerSmith:2004jv, Finkbeiner:2007kk, Arina:2007tm, Chang:2008gd, Cui:2009xq, Fox:2010bu, Lin:2010sb, DeSimone:2010tf, An:2011uq, Pospelov:2013nea,Finkbeiner:2014sja, Dienes:2014via,Bramante:2020zos, Jordan:2018gcd,Bell:2020bes}.  There are interesting and well-motivated scenarios in which the dark sector includes a particle ($\chi_1$) which constitutes the bulk of cosmological dark matter, and a slightly heavier particle ($\chi_2$).  In this case, inelastic scattering of dark matter against Standard Model (SM) particles (that is, $\chi_1~{\rm SM} \rightarrow \chi_2~{\rm SM}$) can produce a subleading population of the heavier particle.  In some scenarios the decay of the heavier particle can in turn produce a photon ($\chi_2 \rightarrow \chi_1 \gamma$), which may be observed as a deposition of electromagnetic energy in a deep underground dark matter detector. This scenario is known as luminous dark matter (LDM)~\cite{Feldstein:2010su}.  Our goal will be to consider this signal in the context of future detectors with excellent energy resolution.

The LDM signal has been considered in previous studies (see, for example,~\cite{Pospelov:2013nea,Eby:2019mgs}). In particular, it was shown that this signal could have potentially explained the excess in electron recoil events seen by 
XENON1T~\cite{Bell:2020bes}. While analyses of the XENONnT data~\cite{Aprile:2022vux} show no such excess, LDM remains a viable scenario.  For non-relativistic dark matter, the decay $\chi_2 \rightarrow \chi_1 \gamma$ produces a nearly monoenergetic photon.  However, since the dark matter is moving relative to the lab frame there is a small width to the photon signal.  A direct detection experiment with sufficiently high energy resolution can resolve the shape of the decay photon energy spectrum. As we will see, information about dark matter particle physics and astrophysics can be unlocked from a detailed analysis of the decay photon spectrum.

In particular, an energy resolution $\sim 10-20 \times$ better than that of XENON1T would allow a direct detection experiment to distinguish between a LDM signal and exothermic electron scattering (depending on the background and exposure of such an experiment).

It has recently been proposed that detectors using diamond as a target material are capable of ${\cal O}$(meV) energy resolutions~\cite{Kurinsky:2019pgb}.  Additionally, in \cite{Griffin:2020lgd}, designs for SiC phonon detectors with similar ${\cal O}$(meV) energy resolution were proposed. We will see that this energy resolution would be sufficient for constraining the velocity dispersion of dark matter with $m_\chi \sim 100~\mev$.  

In the LDM framework, the initial endothermic scatter of the dark matter particle against an SM particle, producing $\chi_2$, need not occur within the detector.  If $\chi_2$ has a very short lifetime, then both the initial scatter (against either nuclei or electrons) and the subsequent decay could deposit energy within the detector.  Otherwise, the initial scatter could occur in the surrounding earth, with the heavier state passing through the detector at the point of decay. In the long lifetime scenario the signal rate therefore scales with the detector volume, rather than the detector mass. This scaling provides a method for discriminating LDM from regular DM scattering, as well as suggesting different detector design optimizations.

The plan of this paper is as follows: in section II we review, in generic terms, the concept of decaying inelastic dark matter and the spectrum of photons it produces. In section III, we consider the ability of current detectors to distinguish 
between different scenarios of new physics which could yield narrow features at $E \sim {\cal O}(\kev)$ in the electron recoil spectrum. In section IV we explore how high resolution detectors can measure the dark matter decay spectrum and deduce its properties. Lastly, in section V we offer some concluding remarks.

\section{Decay spectrum of excited DM}

If dark matter scatters endothermically, it is possible for the heavier state produced by this interaction to decay back to the lighter state.  If this decay proceeds through the production of a photon within the detector, then the decay photon can mimic an electron 
recoil~\cite{Bell:2020bes}.  In the case where dark matter is non-relativistic, then this signal is nearly monoenergetic, with $E \sim \delta \equiv m_{\chi_2} - m_{\chi_1} \ll m_{\chi_1}$.  A similar process where the dark matter directly decays to photons can produce a near monoenergetic peak, smeared by the Doppler effect~\cite{Bessho:2022yyu}.

We  consider the decay process $\chi_2 \rightarrow \chi_1 \gamma$, where we assume that the angular distribution is isotropic, in the rest frame of the parent particle.  This process can arise from a magnetic dipole moment interaction. We can then find the photon spectrum using the results of~\cite{Boddy:2016hbp}.  Although that paper was focused on the indirect detection of dark matter decay, the results are just as relevant here.  Interestingly, these results may, in fact, even be  more useful in the context of the decay of the excited state within a direct detection experiment, since direct detection experiments tend to have better energy resolution than indirect detection experiments.

In the rest frame of $\chi_2$, the energy of the photon is given by 
\bea
E_* &=& \frac{m_{\chi_2}^2 - m_{\chi_1}^2}{2m_{\chi_2}} 
\sim \delta.
\eea 
In the frame of the detector, the photon spectrum is then given by~\cite{Boddy:2016hbp}
\bea 
\frac{dN_\gamma}{dx} &=& \int_{\frac{m_{\chi_2}}{2}
\left(x + \frac{1}{x} \right)}^\infty ~
dE_{\chi_2} \left[ \frac{dN}{dE_{\chi_2}} 
\frac{m_{\chi_2}}{2\sqrt{E_{\chi_2}^2-m_{\chi_2}^2}} 
\right] .
\label{eq:PhotonSpectrum}
\eea 
where $x \equiv E_\gamma / E_*$, and $dN / dE_{\chi_2}$ is the energy spectrum of the $\chi_2$ produced by upscatter.

It is worthwhile to note a few features of 
this spectrum~\cite{Boddy:2016hbp}. First, it is log-symmetric 
about the energy $E_*$, and decreases monotonically as the energy increases or decreases away from this point.  Moreover, the energy spectrum near $E_\gamma = E_*$ is determined by the behavior of $dN/ dE_{\chi_2}$ near $E_{\chi_2} = m_{\chi_2}$ (that is, when the heavier state is produced with very small boost).  In particular, if $dN/ dE_{\chi_2}$ goes to a finite value as  $E_{\chi_2} \rightarrow m_{\chi_2}$, then the photon  spectrum has a sharp spike (the first derivative is discontinuous) at $E_\gamma = E_*$.  If  $dN/ dE_{\chi_2} \rightarrow 0$ at zero boost, then  the photon spectrum has a smooth peak at $E_\gamma = E_*$. But if the $\chi_2$ is only produced with some minimum  non-zero boost (so its injection spectrum vanishes for boosts below some finite value), then the peak of the photon spectrum is actually a flat plateau centered at $E_*$ on a log  scale. We thus see that some qualitative features of dark sector microphysics can be directly related to the decay photon spectrum. 

\begin{figure}[b]
     \centering
     \includegraphics[width=0.9\columnwidth]{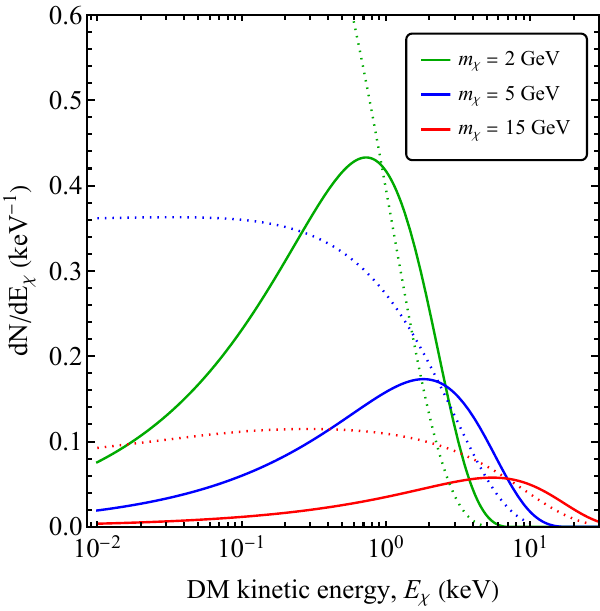}
     \caption{The incoming halo dark matter spectrum (solid) compared to the spectrum of dark matter after inelastic scattering in the Earth (dashed), assuming  $\delta = 1\kev$ and masses: $m_\chi = 2~\gev$ (green), $5~\gev$ (blue) and $15~\gev$ (red).}
     \label{fig:DM_spec}
\end{figure}

We can estimate the energy resolution necessary to exploit these theoretical features of the width of the decay photon spectrum.  Let $\beta = v/c$ describe the rough scale of the speed of $\chi_2$ particles in the laboratory frame.  If a $\chi_2$ moving with speed $\beta$ decays, producing a photon with energy $\sim \delta$ in the $\chi_2$ rest frame, then this photon would have an energy ranging between $\gamma (1-\beta)\delta$ and $\gamma (1+\beta) \delta$ in 
the lab frame.  Since $\beta \ll 1$, we see that the rough width of the dark matter spectrum is $\sim \beta \delta$.  We expect $\beta \lesssim{\cal O}(10^{-3})$.  Moreover, we would need $\delta \lesssim {\cal O}(10^{-6}) m_\chi$ in order for endothermic scattering to be kinematically allowed, for low-mass dark matter.  We thus find that the required energy resolution is ${\cal O}(10^{-9}) m_{\chi_1}$.  

For example, if LDM produced a feature at ${\cal O}(\kev)$ 
in the electron recoil spectrum (which would be observable 
above threshold for Xenon-based detectors), one would need a 
detector with sub-eV energy resolution to probe the shape 
of the energy spectrum.
As another example, we see that for dark matter with a mass of 
$\sim 100 \mev$, one would need an energy resolution of better than $\sim {\cal O}$(100 meV) 
to probe the shape of the recoil spectrum arising from kinematically accessible 
endothermic scattering.  Diamond~\cite{Kurinsky:2019pgb}or SiC~\cite{Griffin:2020lgd} detectors may be produced with meV level resolutions, possibly making either of them a fitting choice.

If the energy resolution of the detector can be ignored, then one can invert Eq.~\eqref{eq:PhotonSpectrum} to obtain 
\bea
\left[ \frac{dN}{dE_{\chi_2}}  
\right]_{E_{\chi_2} = \frac{m_{\chi_2}}{2} \left(x + \frac{1}{x} \right)}  &=& 
\frac{2x}{m_{\chi_2}} {\rm sgn}[1-x] \frac{d^2N_\gamma}{dx^2} .
\label{eq:PhotonSpectrumDerivative}
\eea
In this way a detector with sufficiently fine energy resolution can directly probe the dark matter spectrum.

\section{Luminous Dark Matter with current detectors}

For simplicity, we assume the dark matter scatters off nuclei through an isospin-invariant and spin-independent interaction. Due to the $A^2$ coherent scattering enhancement, we can assume that $\chi_2$ production is dominated by endothermic scattering against relatively heavy nuclei. We also assume that the dark matter is relatively light $m_{\chi_{1,2}} \ll m_A$.  The reduced mass of the dark matter-nucleus system ($\mu_{1,2}$) is thus essentially the same as the dark sector particle mass, and the kinematics of scattering process are independent of the target.  We may thus determine the shape of the $\chi_2$ energy spectrum as a function of $m_{\chi_1}$ and $\delta$, without a detailed assay of the material around the detector (or the entire Earth). Therefore we compute the decay spectra assuming scattering from the Earth's crust only, noting that longer lifetimes which will also upscatter in Earth's core, will not have a significantly altered spectrum. In the appendix, we discuss the effect of the scattering material composition on the energy spectrum and justify these simplifications.
Note that exact composition of the upscattering material does affect the normalization (that is, the flux of $\chi_2$), as does the endothermic scattering cross section, but we will treat this normalization as a free parameter. The normalization is a complex function of the lifetime and cross section but can be computed as in~\cite{Eby:2019mgs}. For simplicity, we assume that the lifetime of $\chi_2$ is long enough that $\chi_2$ will only decay inside the detector if it is produced by scatters which occur outside the detector. For lifetimes $\gtrsim 100$ s, the whole Earth contributes to the observed rate and this scenario can produce detectable rates (at e.g. XENONnT and LZ) for dark matter in the GeV mass range~\cite{Baryakhtar:2020rwy}. 

\begin{figure*}[t]
\centering
     \includegraphics[width=1.8\columnwidth]{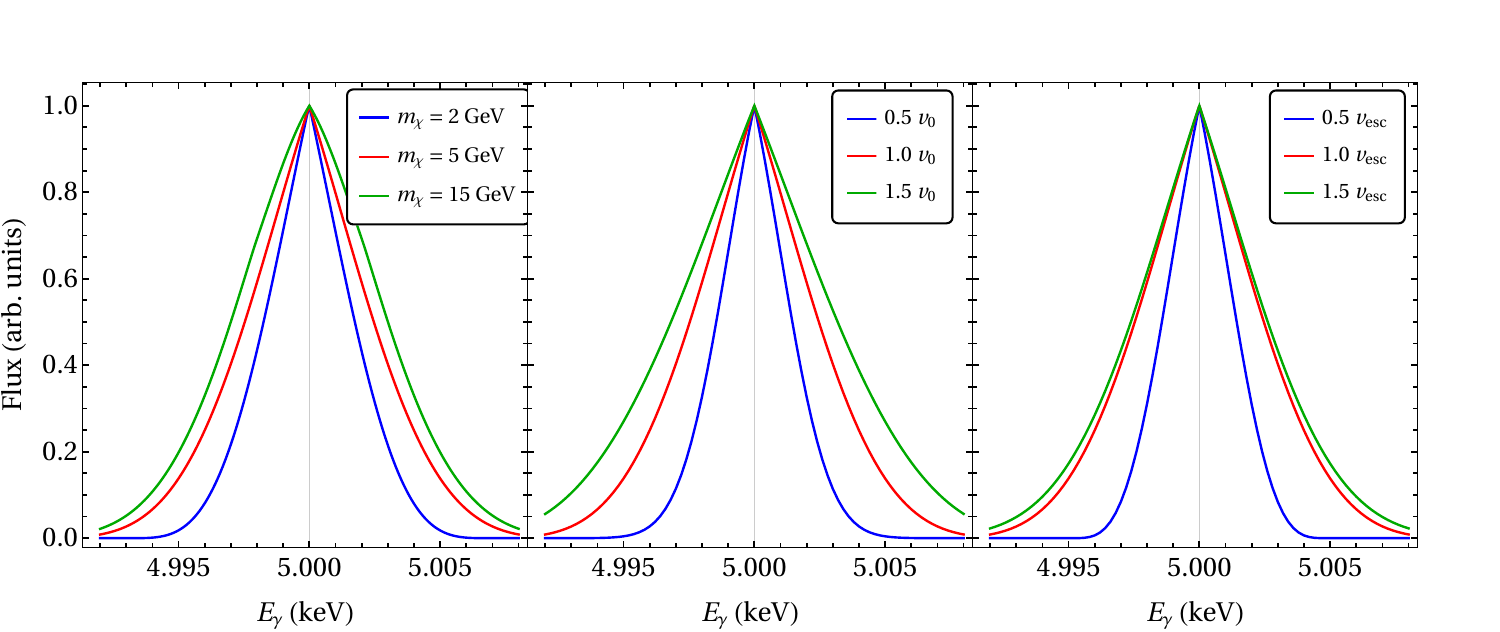}
     \caption{Spectra of photons from the decay of LDM with $\delta=5$ keV varying the DM mass (left), the velocity dispersion (center), and the escape velocity (right). The center and right plots assume a mass of $m_\chi = 5$ GeV.}
     \label{fig:photonSpec}
\end{figure*}

In Figure~\ref{fig:DM_spec}, we plot the spectra of the incoming (solid) and excited (dashed) dark matter particle for three benchmark masses, 
$m_{\chi} = 2~\gev$ (green), $m_{\chi} = 5~\gev$ (blue) and $m_{\chi} = 15~\gev$ (red), with $\delta = 1~\kev$. We have adopted the DM velocity distribution parameters from~\cite{Baxter:2021pqo}: the incoming dark matter particle has a Maxwellian distribution in the frame of the Galactic Center with velocity dispersion $\sigma_0 = v_0/\sqrt{2}$, where $v_0 = 238$~km/s is the local circular velocity. Additionally we take the Sun's peculiar velocity to be $\vec{v}_{\odot,\mathrm{pec}} = (11.1,12.2,7.3)$~km/s and the Earth velocity to be that at March 1st $\vec{v}_e = (29.2,-0.1,5.9)$~km/s, and assume the escape velocity is given by $v_{esc}=544$~km/s.\footnote{Here the vector components are given as $(v_r, v_\phi, v_\theta)$ where $r$ points toward the galactic center and $\phi$ in the direction of the Milky Way's rotation}  

Note that, for all of these benchmarks, the maximum kinetic energy of the outgoing state is 
smaller than of the incoming state, as expected from endothermic scattering.  We also see 
that, although the energy spectrum of the incoming dark particle 
asymptotes to zero at small boost, the energy spectrum of the 
outgoing excited state asymptotes to a finite value at small boost.
This can be understood intuitively.  At the threshold for 
endothermic scattering, the excited state is produced at rest 
in the center-of-mass frame, and thus with small fixed speed in the 
laboratory frame (assuming the dark matter is much lighter than the 
target).  
For incident dark matter with speed slightly 
above threshold, backscattering will then yield an excited state 
at rest in laboratory frame.

In Figure~\ref{fig:photonSpec} (left panel), we plot the photon spectra arising from decay of the excited dark particle 
($\chi_2 \rightarrow \chi_1 \gamma$) for the three benchmark masses.  
In all cases, the photon spectrum exhibits a spike feature 
at $E_\gamma = \delta$, which results from the fact that 
$dN/dE_{\chi_2}$ asymptotes to a finite value at zero boost.  Thus, a tell-tale signature of this 
scenario is that the spectrum has finite width, but still exhibits a cusp.
Note that for all of these benchmark models, the width of the photon spectrum is a few eV.  Thus, if the energy resolution is $\lesssim {\cal O}(\ev)$, one might expect to be able to distinguish these scenarios from that of a true lines signal, and resolve the cusp feature.  
As one can see, a considerably better resolution would be needed to 
distinguish the spectra produced by different dark matter masses from each other.  This is particularly the case when comparing the photon spectra for the cases of relatively heavy dark matter ($m_\chi = 5~\gev$ and $15~\gev$).  
For those cases, the dark matter is heavy enough that endothermic scattering with $\delta = 5~\kev$ causes a relatively small decrease in the kinetic energy of the  dark particle-nucleus system in center-of-mass frame.    Since the kinematics of both cases are similar to that of elastic scattering, 
the velocity distributions of the excited dark particles are the same, resulting in photon signals with 
a width $\sim \beta \delta$ which are similar in both cases.
For $m_{\chi} = 1~\gev$, on the other hand, the kinetic energy of the DM-nucleus system in 
center-of-mass frame is $\sim \delta$.  As a result, the 
outgoing dark particle has a smaller speed relative to the lab frame, yielding a narrower photon 
spectral feature.

To illustrate the effect of the DM velocity distribution on the decay photon spectra, 
we plot a series of benchmark scenarios in Fig.~\ref{fig:photonSpec} (center and right panels).  In 
the center panels, we take the velocity-dispersion to be $0.5~v_0$, $1.0~v_0$, or $1.5~v_0$, 
with the escape velocity taken to be $v_{esc}$.  
In the right panels, we take the escape velocity to be $0.5~v_{esc}$, $1.0~v_{esc}$ or $1.5~v_{esc}$, with the 
velocity dispersion taken to be $v_0$.  In both panels we take $m_\chi = 5~\gev$.

As expected, an increased velocity dispersion leads to a wider 
feature in the photon spectrum.  This results from two effects: 
a larger velocity dispersion increases the typical speed of 
an incoming dark matter particle, and reduces the effect of 
inelasticity on the outgoing particle speed.
Increasing the escape velocity initially has the effect of 
broadening the feature in the photon spectrum, for the same 
reasons.  But after a certain point, these effects 
saturate, because when the escape velocity is much larger than 
the velocity dispersion, the fraction of particles at the 
highest speeds is exponentially small.

\begin{figure}[ht]
     \centering
     \includegraphics[width=0.9\columnwidth]{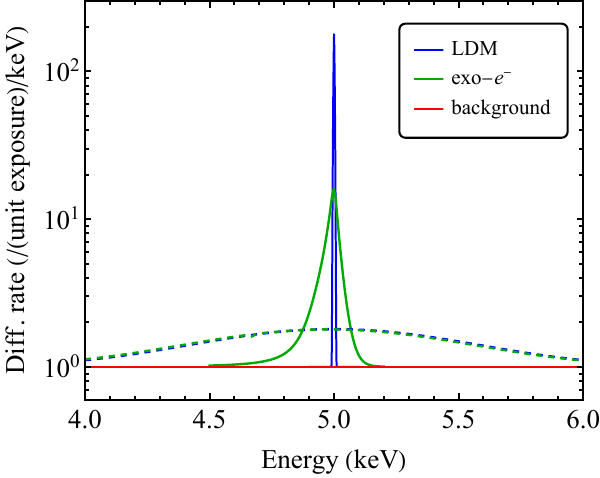}
     \caption{Comparison of the LDM signal with another line-like signal model before (solid) and after (dashed) smearing with a detector resolution of $\sigma=0.5\kev$. Both the LDM spectrum  (blue) and exothermic scattering on electrons spectrum (green) are taken to have $\delta=5\kev$.}
     \label{fig:compSignals}
\end{figure}

We now consider the prospects for a future instrument to 
distinguish between the luminous dark matter scenario and another 
scenario of new physics which would yield a narrow signal. As a benchmark comparison to the LDM model, we will consider the exothermic scattering of dark matter, with a mass of 0.1 GeV, against electrons in a xenon target~\cite{Bramante:2020zos}. When reproducing the calculation for exothermic dark matter scattering we make use of atomic scattering factors from DarkARC~\cite{timon_emken_2019_3581334,Catena:2019gfa}.
The energy resolution of large xenon detectors is $\sim 0.5\kev$ at $E\sim 2-3\kev$ (as can be seen from the argon-37 calibrations in \cite{Aprile:2022vux,LUX-ZEPLIN:2022qhg}). Given this energy resolution, the signals from both the LDM scenario and exothermic scattering against electrons would be indistinguishable from a monoenergetic line. The expected differential event rates for the two models are shown in Figure~\ref{fig:compSignals}.
 
To assess the required number of signal events and resolution required to distinguish these scenarios we generate Asimov datasets assuming an exothermic electron scattering model and try to reject the null hypothesis of a line signal. To do this we use the log-likelihood ratio, $q_\mu$, and calculate the significance as $\sqrt{q_\mu}$~\cite{Cowan:2010js}.  
We adopt, for simplicity, a flat background model which we take to have a signal-to-noise ratio of either 10 or 1 in the region-of-interest. The region-of-interest is taken to be 1 keV either side of the 5 keV peak, split into 40 equal-width bins.

\begin{figure}
     \centering
     \includegraphics[width=0.9\columnwidth]{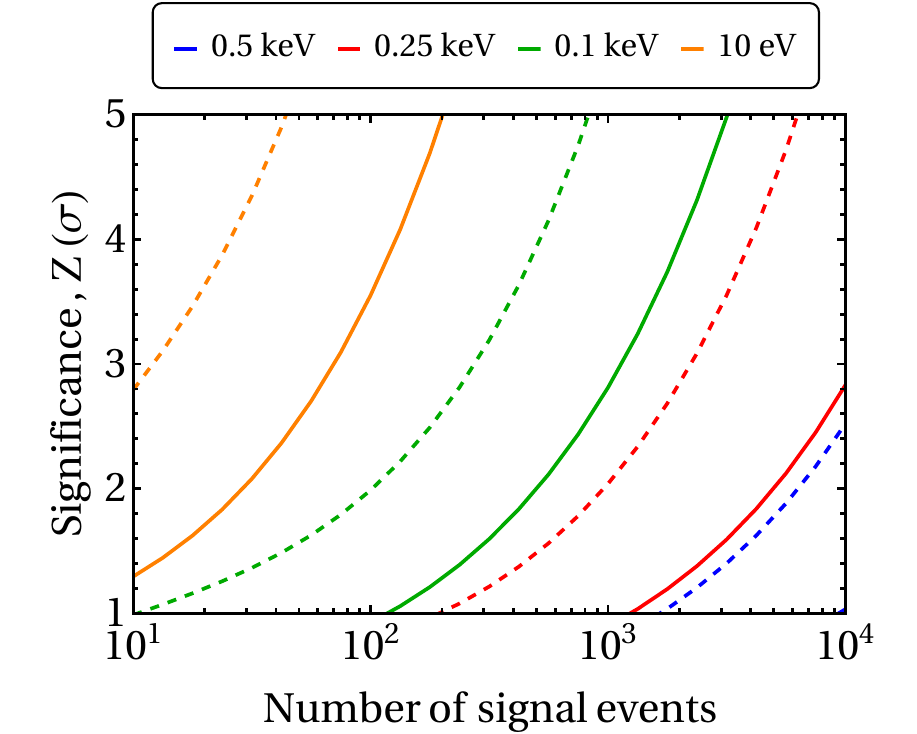}
     \caption{The significance rejecting a line signal model (luminous dark matter) in favor of the exothermic electron scattering model as a function of number of detected events for various resolutions and for a SNR of 0.1 (solid) and 1 (dashed).}
     \label{fig:Significance_Exposure}
\end{figure}

We calculate the significance as a function of the number of signal events for a series of detector resolutions ranging from approximately what can be obtained in a xenon based TPC (0.5 keV) to what can be obtained with current cryogenic detectors (10 eV)~\cite{Ren:2020gaq}. The result is shown in Fig.~\ref{fig:Significance_Exposure}.  
  
As might be expected from Fig.~\ref{fig:compSignals}, an instrument with the energy resolution of a large xenon experiment would need a very large exposure and low background to distinguish between these two scenarios (see the blue curve in Fig.~\ref{fig:Significance_Exposure}).  On the other hand, an improvement in the energy resolution by at least a factor of 5 would allow one to distinguish between these two scenarios at high significance with an exposure that could potentially be realized. 

Note, however, that the situation changes dramatically if one uses multiple detectors with multiple target materials, 
because the luminous dark matter model event rate scales with the volume of the detector, not the target mass.  Thus, 
if signals are seen in two detectors with different target materials, the predicted relative event rates for this scenario of luminous dark matter would be very different from those of a model in which the signal arose from dark matter scattering within the detector.   
This might provide another opportunity for testing this scenario.

Note that, for the energy resolutions which we are considering, the LDM signal is effectively a line signal.  As a result, the analysis does not utilize the detailed shape of the LDM photon signal.  In the next section, we will see that information about the dark matter particle physics and astrophysics can be obtained with better resolution.

\section{Luminous Dark Matter with high-resolution detectors}

We now consider the detection of LDM in a detector with much better resolution than is achievable in large detectors such as liquid noble TPCs. Presently, high resolution detectors are based on solid state technologies that achieve eV-scale resolution~\cite{Kurinsky:2019pgb}. While such detectors typically occupy small volumes, we note that for the LDM scenario one does not require the detector occupy an entire volume. For example, one could surround a vacuum or transparent medium with the high-resolution detectors. For simplicity, in this section we will not consider specific detector designs or configurations and instead consider benchmarks based on resolution and background (parameterized as the signal-to-noise ratio, or SNR, in the region-of-interest). We will assume the background is flat in energy within the region of interest - a good approximation given the small width of the LDM signal. See Table~\ref{tab:dets} for the detector parameters chosen as benchmarks. 

\begin{table}[bth]
    \centering
    \caption{Benchmark detector parameters}
    \begin{tabular}{ccc}
    \hline
    \hline
         & Standard & High-performance\\
         \hline
         resolution & 1 eV & 2 meV\\
         background (SNR) & 0.1 & 1\\
    \hline
    \hline
    \end{tabular}
    \label{tab:dets}
\end{table}

We will consider three LDM benchmark models with $\delta=5,\,1\mathrm{\,\,and\,\,} 0.1\kev$ and $m_{\chi_1}=5,\,\,2,\,\,1\gev$ respectively. Rather than select a cross section, lifetime and detector volume, we choose to explore the prospects of detection and reconstruction of parameters using the number of photons detected. For the chosen values of $\delta$ we optimistically assume that $\mathcal{O}(10^3)$ events could be obtained. In the $\delta=5\kev$ case, obtaining $\mathcal{O}(10^3)$ events would require a very large volume detector, considering current constraints on the event rate from XENONnT. For the lower mass splitting cases, which is not currently strongly constrained, technological improvements and novel detector designs could rapidly make this a possibility.

\begin{figure}
     \centering
     \includegraphics[width=0.9\columnwidth]{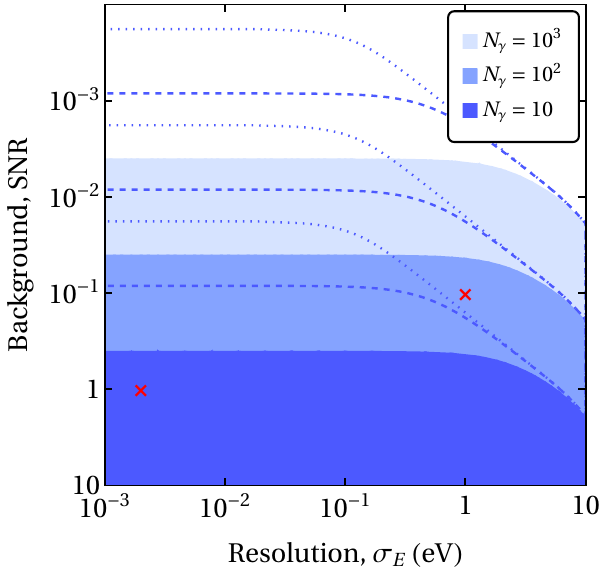}
     \caption{Regions where $3\sigma$ local significance can be obtained as a function of the background and resolution, for three different exposure levels (see legend) and for $\delta=5\kev$ (solid) and $\delta=1\kev$ (dashed) and $\delta=0.1\kev$ (dotted). The standard and high-performance detector benchmarks are indicated with red crosses.}
     \label{fig:Significance_Resolution}
\end{figure}

To assess the sensitivity of future high-resolution detectors to LDM signals we first compute the regions of detector resolution vs. SNR that would admit a $3\sigma$ (local significance) detection for various numbers of signal events. This is performed by calculating the $\Delta\chi^2$ of the signal + background vs. the background-only hypothesis in a 20 eV window around the spectral peak, partitioned into 100 bins. The result is shown in Fig.~\ref{fig:Significance_Resolution}. For detector resolutions down to around $1 \ev$ we see a large improvement in a detector's ability to pick the signal out from the background. However, with the characteristic width of the LDM spectrum being $\sim\beta\delta$ we see that resolutions below $\sim1\ev$ offer diminishing improvement. This also explains why sensitivity to the $\delta=0.1$ and $1\kev$ signals benefit more for smaller resolutions than in the case of the $\delta=5\kev$ signal.

If a detection is made and the shape of the spectrum is measured to high enough precision we can then perform inference on the LDM parameters. Clearly the value of $\delta$ will be known to high precision, but the other parameters may not be probed as easily.

\begin{figure*}
      \centering
      \includegraphics[width=2\columnwidth]{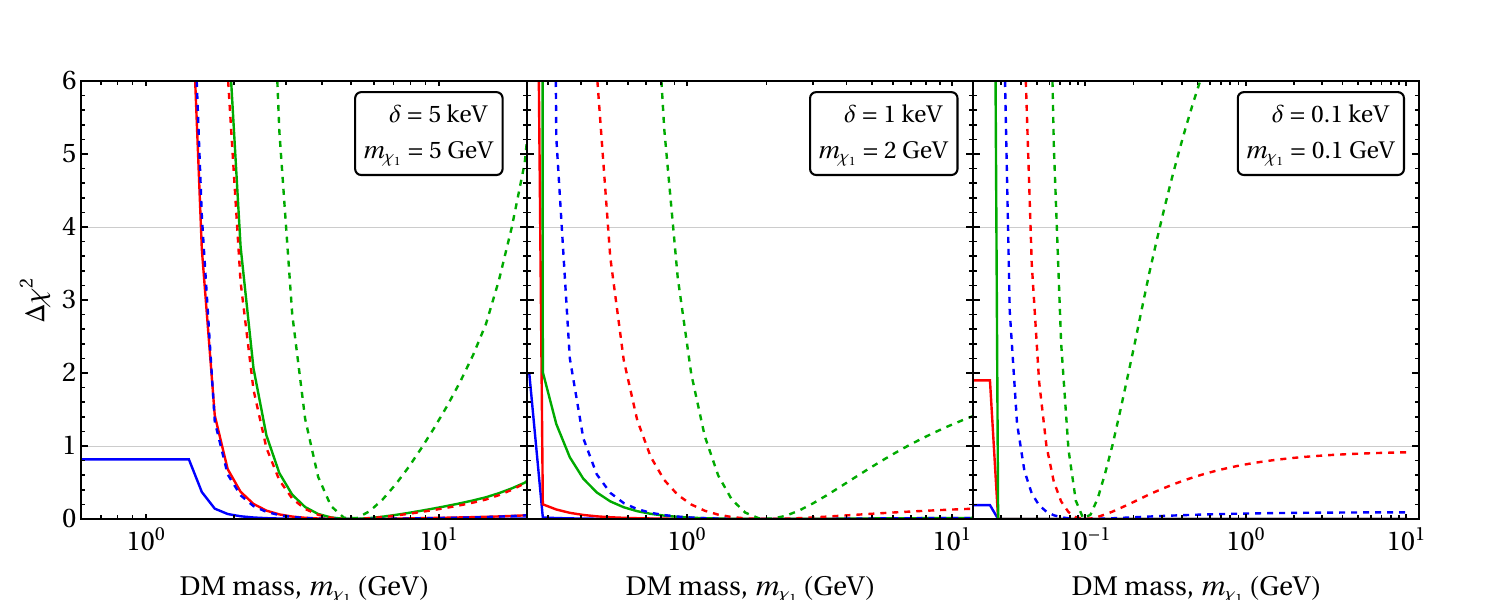}\\
      \caption{The $\Delta \chi^2$ as a function of mass with  $\delta=5$ keV and $m_\chi = 5~\gev$ (left), $\delta=1$ keV and $m_\chi = 2~\gev$ (center) and $\delta=0.1$ keV and $m_\chi =0.1    ~\gev$ (right) for standard (solid) and high-performance (dashed) detector assumptions and three different numbers of detected events 10 (blue), 100 (red) and 1,000 (green).}
   \label{fig:bestFitMass}
 \end{figure*}

In Figure~\ref{fig:bestFitMass}, we plot $\Delta \chi^2$ as a 
function of mass for three scenarios of the true model 
parameters: $m_{\chi_1} = 5~\gev$, $\delta = 5~\kev$ (left),
$m_{\chi_1} = 2~\gev$, $\delta = 1~\kev$ (center), 
and $m_{\chi_1} = 0.1~\gev$, $\delta = 0.1~\kev$ (right).  
In each case, we assume either a standard performance (solid) 
or high-performance (dashed) detector, and either 
10 (blue), 100 (red) or 1000 (green) signal events detected. 
As anticipated in Figure~\ref{fig:photonSpec}, when the true 
model has $m_{\chi_1} = 5~\gev$, $\delta = 5~\kev$ (left), 
a larger hypothesized mass will yield a nearly identical 
spectrum, which is difficult to reject even with many events, 
and a high-performance detector.  But a smaller hypothesized 
mass will be distinguishable at the eV scale, and can be 
rejected even with a standard performance detector.  
But for $\delta \ll {\cal O}(\kev)$, the width of the 
spectrum will be $\ll \ev$, and a high-performance detector will 
be needed to distinguish the particle mass.

\begin{figure*}
     \centering
     \includegraphics[width=2\columnwidth]{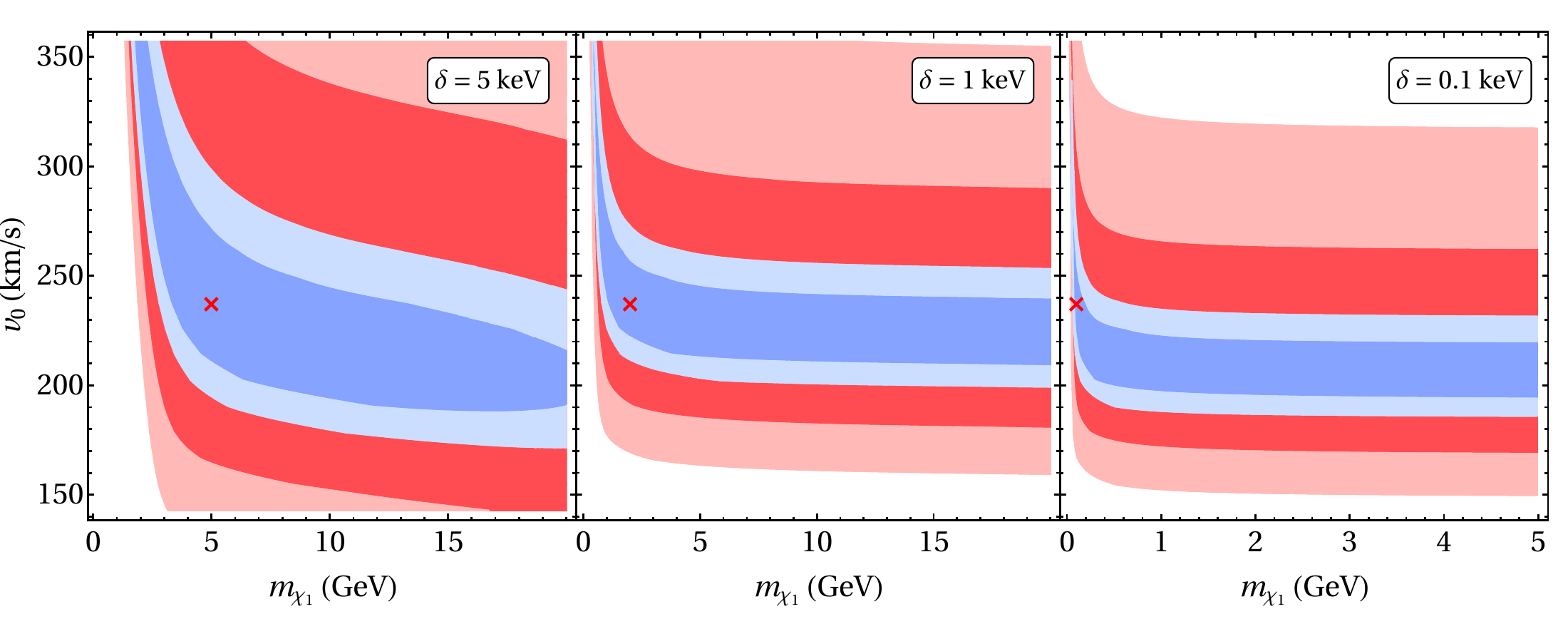} \\
     \includegraphics[width=2\columnwidth]{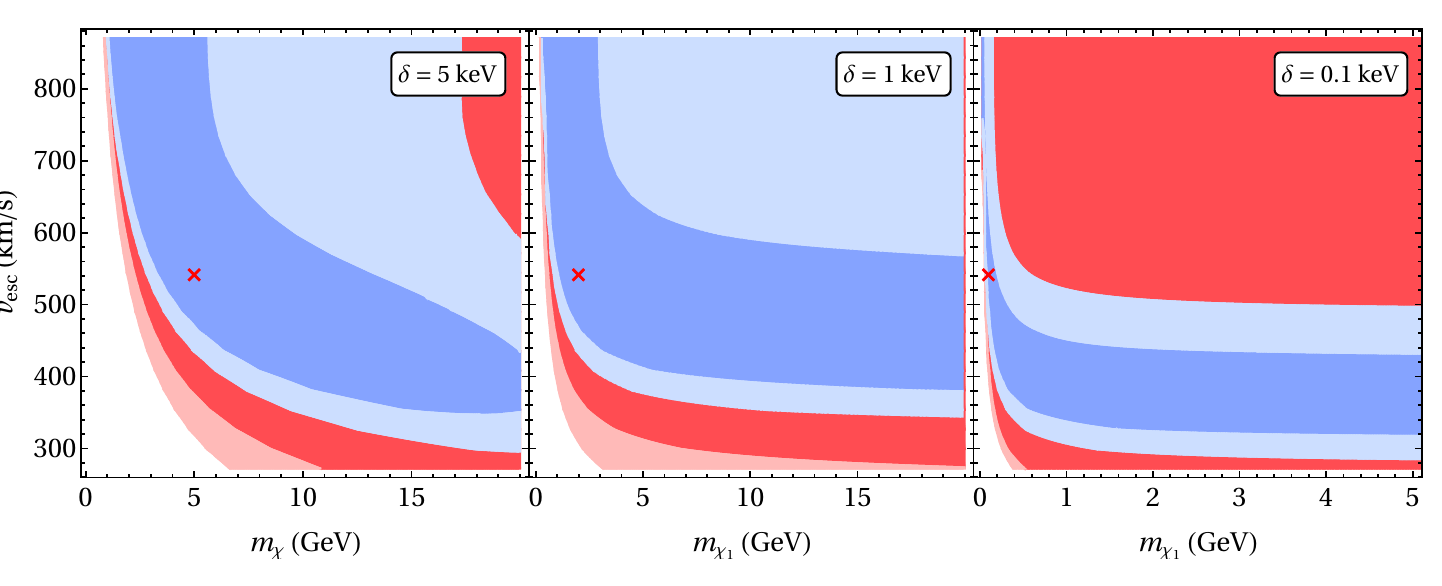}\\
     \caption{1$\sigma$ (dark) and 2$\sigma$ (light) contours in $m_\chi$ vs. $v_0$ (top) and $m_{\chi}$ vs. $v_{\mathrm{esc}}$ (bottom) for 100 (red) and 1000 (blue) events with the high-performance detector.}
     \label{fig:velInf}
\end{figure*}

In Figure~\ref{fig:velInf}, we plot $1\sigma$ (dark) and 
$2\sigma$ (light) parameter constraints in the $(m_\chi , v_0)$ 
(top panels) and $(m_\chi, v_{esc})$ (bottom panels) planes, assuming a high-performance detector 
and either 100 (red) or 1000 (blue) signal events observed.  
The true model assumes $v_{esc}=544$ km/s, $v_0 = 238$ km/s, 
and either 
$m_{\chi_1} =5~\gev$, $\delta = 5~\kev$ (left panels),
$m_{\chi_1} =2~\gev$, $\delta = 1~\kev$ (center panels),
or $m_{\chi_1} =1~\gev$, $\delta = 0.1~\kev$ (right panels).
In each panel, the true model is denoted with a red cross.   

We see that one has the  ability to reconstruct the velocity dispersion, with enough events.
But for a mass hypothesis which is smaller than the true mass, the reconstructed velocity dispersion tends 
to lie above the true velocity dispersion.  In this case, the mass hypothesis leads to  
endothermic scattering which is closer to threshold, yielding a narrower signal.  This effect 
is compensated by increasing the velocity dispersion.

On the other hand, it is difficult to reject any reasonable hypothesis for the escape 
velocity, largely because of the small fraction of events which lie at the tail 
of the velocity distribution.  Note, however, that we are considering here benchmark 
scenarios for which a large fraction of the dark matter is above threshold for endothermic 
scattering.  For scenarios in which only particles on the tail of the velocity distribution 
are above threshold, we would expect a much stronger ability to constrain the escape 
velocity.

\section{conclusion} \label{sec:conclusion}

We have considered the (in)direct detection of luminous dark matter (LDM) with 
detectors with good energy resolution.  In this scenario, dark matter scatters 
endothermically ($\chi_1 A \rightarrow \chi_2 A$) with either the detector material or the surrounding earth, producing 
a slightly heavier particle.  This excited particle then decays ($\chi_2 \rightarrow \chi_1 \gamma$) 
within the volume of a direct detection experiment, producing a photon which yields 
an electron recoil signal.  In this case, the direct detection experiment actually functions 
as an indirect detection experiment, measuring not the energy deposited by the scattering 
of dark matter against the target, but the energy of the photon produced by dark particle decay.
We have found that, with improved energy resolution, one can probe the spectral features of 
the photon signal, allowing one to reconstruct information about dark matter particle physics 
and astrophysics.

For example, with an order of magnitude improvement in the energy resolution beyond that obtained by current xenon detectors, one can distinguish this LDM scenario from other scenarios of beyond-the-Standard-Model physics yielding an narrow electron recoil signal 
in the ${\cal O}(\kev)$ range. Moreover, the detailed shape of the photon spectrum also carries information about the dark matter 
velocity distribution.  A high-performance detector (with specifications comparable to detectors 
under development) would be able to reconstruct some parameters of the velocity distribution 
(such as the velocity dispersion), though other parameters (such as the escape velocity) are 
far more challenging.

Interestingly, if the lifetime of the excited state is sufficiently long, the excited states 
which decay inside the detector would originate in endothermic scattering events in the earth 
outside the detector.  In this case, the event rate at any detector would scale as the fiducial 
volume, not the fiducial mass.  One could increase the fiducial volume of a high-performance 
detector without increasing the instrumented mass by having the detector material enclosed a 
volume of vacuum or transparent medium.  This type of detector presents a variety of opportunities 
and technical challenges which would be interesting to explore in future work.

\begin{acknowledgments}
The work of BD is supported in part by the DOE Grant No. DE-SC0010813. The work of JK is supported in part by DOE grant DE-SC0010504. The work of JLN and NFB is supported by the Australian Research Council through the ARC Centre of Excellence for Dark Matter Particle Physics, CE200100008. JBD acknowledges support from the National Science Foundation under grant no. PHY-2112799. 

\end{acknowledgments}

\appendix
\section{Dark matter upscattering in the Earth}

Ignoring considerations of the overall normalization, the spectrum of DM scattered from a nuclear target $A$ is
\begin{eqnarray}
    \frac{dN^i_\chi}{dE_\chi} & \propto & \frac{m_A}{2 m_{\chi_1} \mu_{\chi p}^2} A^2 \int_{v>v_{\mathrm{min}}} F^2(q^2) \frac{f(v)}{v} dv
    \\
    &\approx&  \frac{m_A}{2 m_{\chi_1} \mu_{\chi p}^2} A^2 G(v_{\mathrm{min} })
\end{eqnarray}
where the approximation denotes taking the low-momentum transfer limit (i.e. ignoring the effect of the nuclear form factor, $F^2(q^2)$, which allows for a significant numerical simplification) and $v_{min}$ is the minimum speed such that inelastic scattering can yield an outgoing particle with energy $E_\chi$.

The total spectrum is found by summing over the elemental scattering targets: 
\begin{equation}
    \frac{dN_\chi}{dE_\chi} = \sum_{i} \frac{n_i}{n_{tot}} \frac{dN^i_\chi}{dE_\chi}
\end{equation}
where $n_i$ is the number density of the $i$th isotope, and $n_{tot} = \sum_i n_i$ is the total number density. Note that, beyond an overall scaling, the only dependence of the shape of the energy spectrum on the $n_i$ arises from $v_{min}$, which depends on the nucleus mass.

\begin{figure}
    \centering
    \includegraphics[width=.9\columnwidth]{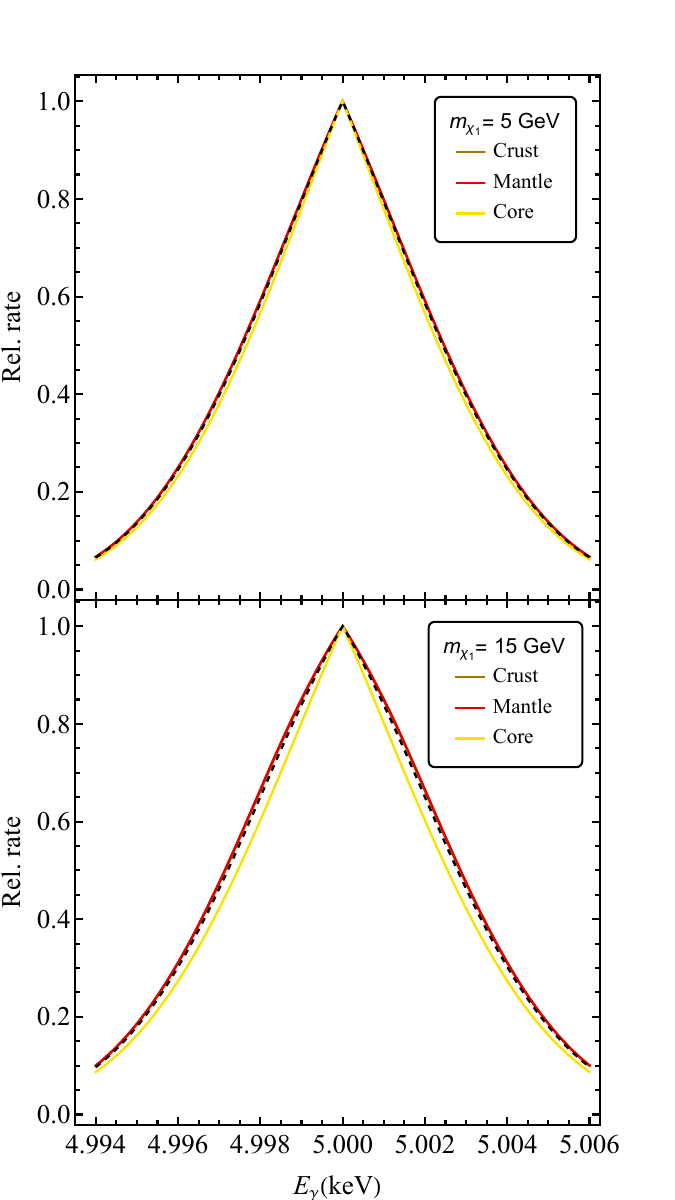}
    \caption{Photon spectra arising from decaying DM after upscattering in different regions of the Earth, assuming 
    $\delta = 5~\kev$ and either $m_{\chi_1} = 5~\gev$ (top) or $m_{\chi_1} = 15~\gev$ (bottom) (note the crust and mantle curves are degenerate). The dashed curves represent a weighted average of contributions from the three regions that applies in the long-lifetime limit.}
    \label{fig:DM_spec_comp}
\end{figure}

In general the $n_i$ will be a function of the $\chi_2$'s lifetime, which dictates how far from the scattering location the $\chi_2$ can travel to reach the detector. In the short lifetime limit this includes only the crust surrounding the detector and in the long lifetime limit the whole Earth would contribute. To demonstrate the insensitivity of the photon spectrum to the precise composition of the scattering targets we show in Fig.~\ref{fig:DM_spec_comp} the resulting photon spectra, assuming $\delta  = 5\kev$ and either 
$m_{\chi_1} = 5\gev$ (top panel) or $m_{\chi_1} = 15\gev$ (bottom panel), and assuming the crust's composition (used in the main analysis) vs.~the mantle and core's composition (using data from \cite{MCDONOUGH2014559}). 
The worst-case scenario can be obtained assuming the long-lifetime limit (shown as the dashed curve in Fig.~\ref{fig:DM_spec_comp}). We see that, even for $m_{\chi_1} = 15\gev$, the effect of the Earth's composition (and thus DM lifetime) is very small and is subdominant to the changes due to the mass and velocity distribution of the DM.

\bibliographystyle{apsrev4-1.bst}
\bibliography{migdal.bib}

\end{document}